# Reflection on modern methods: a note on variance estimation when using inverse probability weighting to handle attrition in cohort studies

Marie-Astrid METTEN, Univ Rennes, CHU Rennes, Inserm, EHESP, Irset (Institut de recherche en santé, environnement et travail) - UMR_S 1085, F-35000 Rennes, France

Nathalie COSTET, Univ Rennes, Inserm, EHESP, Irset (Institut de recherche en santé, environnement et travail)-UMR_S 1085, Rennes, France

Jean-François VIEL, Univ Rennes, CHU Rennes, Inserm, EHESP, Irset (Institut de recherche en santé, environnement et travail) - UMR_S 1085, F-35000 Rennes, France

Guillaume CHAUVET, Univ Rennes, ENSAI, CNRS, IRMAR - UMR 6625, F-35000 Rennes, France

**Abstract**

The inverse probability weighting (IPW) method is used to handle attrition in association analyses derived from cohort studies. It consists in weighting the respondents at a given follow-up by their inverse probability to participate. Weights are estimated first and then used in a weighted association model. When the IPW method is used, instead of using a so-called naïve variance estimator, the literature recommends using a robust variance estimator. However, the latter may overestimate the variance because the weights are considered known rather than estimated. In this note, we develop, by a linearization technique, an estimator accounting for the weight estimation phase and explain how it differs from naïve and robust variance estimators. We compare the three variance estimators through simulations under several MAR and MNAR scenarios. We found that both the robust and linearized variance estimators were approximately unbiased, even in MNAR scenarios. The naive variance estimator severely underestimated the variance. We encourage researchers to be careful with variance estimation when using the IPW method, avoiding naïve estimator and opting for a robust or linearized estimator. R and SAS codes are provided to implement them in their own studies.

**Key messages**

- The inverse probability weighting (IPW) method is used to handle attrition in cohort studies.
- The true probability of response is unknown and estimated from the data using a so-called response model.
- A robust variance estimator is recommended, but it may overestimate the variance as weights are considered known rather than estimated.
- We propose a linearized variance estimator accounting for the weight estimation phase and show it is approximately unbiased in all MAR and MNAR scenarios tested.

**Introduction**

Cohort studies are largely favored by the epidemiologists to investigate causal associations between exposure and health outcomes. However, due to their longitudinal design, these studies often suffer from increasing attrition (non-participation, withdrawal) of the participants along the successive follow-ups. Attrition results in missing data for all the variables collected at a follow-up (including the health outcome of interest) for a more or less large subgroup of the initial participants.

The simplest and most widely used approach to handle attrition in cohort studies is complete-case analysis (CCA). It consists in conducting the analysis within the subset of the respondents at the follow-up only. This method assumes a *missing completely at random* (MCAR) attrition mechanism[1] where the probability of attrition data does not depend on either the values of previously observed variables or the values of the missing variables at the follow-up. However, in real life, other attrition mechanisms are more likely to happen, such as a *missing at random* (MAR) or *missing not at random* (MNAR) mechanisms. In the MAR situation, the probability of attrition data depends on the values of observed variables but not on the values of the missing variables at the follow-up. In MNAR situation, the probability of attrition data depends on the values of the missing variables at the follow-up.

In the epidemiological literature, some methods have been proposed to handle attrition, among them the inverse probability weighting (IPW) method.[2,3] The principle is to redress the sample of respondents with weights reflecting their probability to participate at the follow-up, in order to recreate the initial cohort. Weights are defined as the inverse of the probability to respond, so that respondents with a lower probability of response are given a higher weight in the analysis.

The true probability of response is unknown and needs to be estimated from the data using a model (the response model) to obtain the weights that will be used in a second step for the modeling of the association of interest (the association model). [4–6] Particular care should be taken for the method of estimating variance when using the IPW method. The simplest option is to use a naive variance estimator, which is often given by default in statistical softwares, without further specification. As an alternative, several methodological publications have recommended using a robust or "sandwich" variance estimator.[7–9] However, robust estimators tend to be conservative because they assume that weights are known rather than estimated.[7,10] Intuitively, the use of estimated weights leads to lower variance because the weights account for the observed response probability, rather than the expected response probability. In the applied literature, the use of a robust variance estimator is sometimes mentioned[11,12] but, in most cases, the way in which the variance was estimated is not mentioned, suggesting the use of naive variance estimators.

In this article, we develop a variance estimator that accounts for the fact that response probabilities are estimated rather than known with certainty using the linearization technique described by Deville[13] (called in the rest of the paper linearized estimator). We explain how it differs from the more widely used naive and robust variance estimators, and we evaluate all three through a simulation study under several scenarios of attrition mechanisms.

**1. Estimating variance using inverse probability weighting**

We first introduce the response model used to obtain the estimated response probabilities in Section 1.1. In Section 1.2, we consider the case in which the association model can be described by a linear regression and we present the options for the estimation of the variance of coefficients from weighted association models.

*1.1 Estimation of the response probabilities*

We note $R_i$ as a response indicator, equal to 1 if the individual *i* responds and 0 otherwise. Let $X_i$ be a vector of covariates fully observed in the sample. We note $p_{Ri}^* = \Pr(R_i = 1|X_i)$. It is assumed that such probabilities are given by a logistic regression model:

$$\text{logit}(p_{Ri}^*) = X_i^T \alpha_* \Leftrightarrow p_{Ri}^* = \frac{\exp(X_i^T \alpha_*)}{1 + \exp(X_i^T \alpha_*)} \qquad (1)$$

Let $\hat{\alpha}$ be the estimator of the coefficient $\alpha_*$, obtained by solving the estimating equation of the logistic regression:

$$\sum_{i=1}^{n} X_i \left\{ R_i - \frac{\exp(X_i^T \alpha)}{1 + \exp(X_i^T \alpha)} \right\} = 0 \qquad (2)$$

We let $\hat{p}_{Ri}$ denote the estimator of the response probability, obtained by replacing the coefficient $\alpha_*$ in (1) by its estimator $\hat{\alpha}$. After computation, we obtain the approximate expression for the individual weights:

$$\frac{1}{\hat{p}_{Ri}} \simeq \frac{1}{p_{Ri}^*} - \left( \frac{1}{p_{Ri}^*} - 1 \right) X_i^T \left\{ \sum_{j=1}^{n} p_{Rj}^*(1 - p_{Rj}^*) X_j X_j^T \right\} \sum_{j=1}^{n} (R_j - p_{Rj}^*) X_j \qquad (3)$$

*1.2 Estimation of the coefficients and their variance in a weighted linear regression (association model)*

In this Section, it is assumed that the outcome of interest $Y_i$ is related to a vector $Z_i$ of covariates, according to the linear regression model (association model):

$$m: Y_i = Z_i^T \beta_* + \epsilon_i \text{ with } \begin{cases} E_m(\epsilon_i) = 0, \\ V_m(\epsilon_i) = \sigma_*^2 v_i. \end{cases} \qquad (4)$$

The coefficients $\beta_*$ and $\sigma_*$ are not known but the variable $v_i$ giving the individual variance structure is assumed to be known.

Let $\hat{\beta}$ denote the estimator of the coefficient $\beta_*$, obtained by solving the weighted estimating equation of the linear regression:

$$\sum_{i=1}^{n} \frac{R_i}{\hat{p}_{Ri}} v_i^{-1} Z_i \{Y_i - Z_i^T \beta\} = 0. \qquad (5)$$

**1.2.1 Naïve estimator**

The variance is estimated as if the specified weights were the inverse of the individual variances $v_i$ in the association model given in Eq 4. This leads to the naïve variance estimator:

$$\hat{V}_{naive}(\hat{\beta}) = \left( \sum_{i=1}^{n} \frac{Z_i Z_i^T}{\hat{p}_{Ri}} \right)^{-1}. \tag{6}$$

The naïve estimator does not make use of the correct variance-covariance structure of the association model and may therefore be severely biased.

### 1.2.2 Linearized estimator

We can rewrite equation (5) as:

$$\sum_{i=1}^{n} \frac{R_i}{\hat{p}_{Ri}} v_i^{-1} Z_i \{Y_i - Z_i^T \beta_*\} = \left\{ \sum_{i=1}^{n} \frac{R_i}{\hat{p}_{Ri}} v_i^{-1} Z_i Z_i^T \right\} \{\hat{\beta} - \beta_*\},$$
$$\simeq \left\{ \sum_{i=1}^{n} \frac{R_i}{p_{Ri}^*} v_i^{-1} Z_i Z_i^T \right\} \{\hat{\beta} - \beta_*\}. \tag{7}$$

The second (approximate) equality comes from the fact that $\hat{\beta} - \beta_*$ is an $o_p(1)$ and thus the error by replacing $\hat{p}_{Ri}$ by $p_{Ri}^*$ is negligible.

By plugging in the approximate expression obtained in (3) into the first member of (7), we obtain:

$$\sum_{i=1}^{n} \frac{R_i}{\hat{p}_{Ri}} v_i^{-1} Z_i \{Y_i - Z_i^T \beta_*\} \simeq \sum_{i=1}^{n} \left[ \frac{R_i}{p_{Ri}^*} v_i^{-1} Z_i \{Y_i - Z_i^T \beta_*\} - \{R_i - p_{Ri}^*\} \gamma_*^T X_i \right], \tag{8}$$

with:

$$\gamma_* = \left\{ \sum_{j=1}^{n} p_{Rj}^*(1 - p_{Rj}^*) X_j X_j^T \right\}^{-1} \sum_{j=1}^{n} R_j \left( \frac{1}{p_{Rj}^*} - 1 \right) v_j^{-1} \{Y_j - Z_j^T \beta_*\} X_j Z_j^T. \tag{9}$$

From equation (7) it follows that $\hat{\beta} - \beta_* \simeq \sum_{i=1}^{n} U_i$, with:

$$U_i = \left\{ \sum_{j=1}^{n} \frac{R_j}{\hat{p}_{Rj}} v_j^{-1} Z_j Z_j^T \right\}^{-1} \left\{ \frac{R_i}{\hat{p}_{Ri}} v_i^{-1} Z_i \{Y_i - Z_i^T \hat{\beta}\} - \{R_i - \hat{p}_{Ri}\} \hat{\gamma}^T X_i \right\}, \quad (10)$$

and with:

$$\hat{\gamma} = \left\{ \sum_{j=1}^{n} \hat{p}_{Rj}(1 - \hat{p}_{Rj}) X_j X_j^T \right\}^{-1} \sum_{j=1}^{n} R_j \left( \frac{1}{\hat{p}_{Rj}} - 1 \right) v_j^{-1} \{Y_j - Z_j^T \hat{\beta}\} X_j Z_j^T. \quad (11)$$

The linearized estimator of the variance-covariance matrix $V(\hat{\beta})$ is:

$$\hat{V}_{lin}(\hat{\beta}) = \frac{n}{n-1} \sum_{i=1}^{n} U_i U_i^T. \quad (12)$$

Similar developments for cases of logistic regression are provided in the Supplementary Appendix 1.

### 1.2.3 Robust estimator

Alternatively, the so-called robust variance estimator is obtained as follows: in the definition of the linearized variable $U_i$ given in equation (10), the component $-\{R_i - \hat{p}_{Ri}\}\hat{\gamma}^T X_i$ accounts for the estimation of the response probabilities. Suppressing this term is equivalent to considering that the response probabilities are known rather than estimated. This leads to:

$$\hat{V}_{rob}(\hat{\beta}) = \frac{n}{n-1} \sum_{i=1}^{n} V_i V_i^T, \quad (13)$$

where:

$$V_i = \left\{ \sum_{j=1}^{n} \frac{R_j}{\hat{p}_{Rj}} v_j^{-1} Z_j Z_j^T \right\}^{-1} \left\{ \frac{R_i}{\hat{p}_{Ri}} v_i^{-1} Z_i \{Y_i - Z_i^T \hat{\beta}\} \right\}. \quad (14)$$

If the response probabilities $p_{Ri}$ were known, this variance estimator would be approximately unbiased, even if the association model was mis-specified, hence its name. The R and SAS codes to compute robust and linearized variance estimators are given in the Supplementary Appendix 2.

## 2. Simulation study

We conducted a Monte-Carlo simulation under several MAR and MNAR scenarios to compare the performance of the three variance estimators (naïve, robust, linearized). We focused on the case of a linear regression model in which a continuous outcome is explained by continuous exposure and covariates.

### 2.1 Data-generating process

We compared the performance of the three variance estimators under various response model specifications, differing according to the role played by the variables (predictor of the outcome, of the exposure, or both outcome or exposure …).. We first created a sample of size $n = 1,000$, containing seven covariates $z_1, …, z_7$ generated independently according to standard normal distributions. We then generated an exposure variable according to the following model:

$$x_i = 1 + \alpha_1 z_{1i} + \alpha_2 z_{2i} + \alpha_5 z_{5i} + \alpha_6 z_{6i} + \epsilon_i, \qquad (15)$$

where $\epsilon_i$ is generated according to a standard normal distribution. In the exposure model (15), the coefficients were chosen such that the correlation between $x_i$ and each of the covariates $z_i$ was approximately 0.2. We generated an outcome variable according to the following model:

$$y_i = 1 + \beta x_i + \beta_1 z_{1i} + \beta_3 z_{3i} + \beta_5 z_{5i} + \beta_7 z_{7i} + \epsilon'_i, \qquad (16)$$

where $\epsilon'_i$ is generated according to a standard normal distribution. In the outcome model (16), the coefficients were chosen such that the correlation between $y_i$ and $x_i$ was approximately 0.3, and the correlation between $y_i$ and any of the covariates $z_i$ was approximately 0.2. Finally, we generated response probabilities according to the following logistic model:

$$\text{logit}(p_i) = \gamma_0 + \gamma_y\, y_i + \gamma_x\, x_i + \gamma_1\, z_{1i} + \gamma_2\, z_{2i} + \gamma_3\, z_{3i} + \gamma_4\, z_{4i}, \tag{17}$$

We used the values $\gamma_y = 0.0, 0.2$ or $0.5$ and $\gamma_x = 0.0, 0.2$ or $0.5$. The case in which $\gamma_y$=0.0 corresponds to a MAR situation. The cases in which $\gamma_y$=0.2 and 0.5 correspond to MNAR situations. In the response model (17), the coefficients $\gamma_1, \gamma_2, \gamma_3$, and $\gamma_4$ were chosen to be equal to 0.1. The coefficient $\gamma_0$ was chosen such that the average response rate was approximately 60% for all cases. In the sample, the individuals responded independently with the probabilities $p_i$. The data-generation model is presented in Figure 1 and the nine response mechanism scenarios are summarized in Table 1.

*2.2 Simulation performance criteria*

The performance of the naïve, robust, and linearized variance estimators was assessed using the response and association models that perfectly matched the data generation models.

We performed 10,000 simulations per scenario. The results are expressed in terms of Monte Carlo relative bias, namely:

$$RB_{MC}(\hat{V}(\hat{\beta})) = \frac{B_{MC}(\hat{V}(\hat{\beta}))}{V(\hat{\beta})} \text{ with } B_{MC}(\hat{V}(\hat{\beta})) = \frac{1}{10{,}000} \sum_{b=1}^{10{,}000} \left(\widehat{V^b}(\hat{\beta}) - V(\hat{\beta})\right),$$

and where the variance $V(\hat{\beta})$ was obtained through an independent simulation run of 10,000 simulations. The simulations were conducted using SAS version 9.4.

**3. Results of the simulation study**

We plotted the relative bias of each variance estimator for the nine response mechanism scenarios (Figure 2). The variance was significantly underestimated by the naive estimator in all scenarios. This phenomenon was even more pronounced in MNAR scenarios 3, 4, 5, and 6.

The performance of the robust and linearized estimators was similar. They correctly estimated the variance for all scenarios, including MNAR scenarios.

## 4. Discussion

Several authors have proposed the IPW method to handle attrition in cohort studies.[2,3] The care that must be taken with variance estimation when using this method is relatively unknown to applied researchers.

Our simulation study compared the performance of two variance estimators possibly used in common practice (naïve and robust estimators) with that of a linearized estimator that accounts for the weight estimation phase. We showed that the use of naïve variance estimators should be avoided because they strongly underestimate variance. The linearized variance estimator, for which we presented the explicit formulas, estimated the variance approximately unbiasedly, even in MNAR scenarios. Publications within the inverse-probability-of-treatment weighting (IPTW) framework (propensity score) also showed the variance of the treatment effect to be accurately estimated when the weight estimation phase was included in the calculation of the variance estimator.[14–17]

We expected an overestimation of the variance with the robust variance estimator, which does not account for the weight estimation phase. However, our simulation study showed that the results obtained using robust and linearized estimators were similar in all scenarios. Enders et al. also reported negligible differences between results obtained using a robust estimator and an estimator that accounts for the weight estimation phase for the IPTW method in survival analysis and ultimately recommended using the robust estimator for practical reasons (better implementation in software).[18] In our case, researchers can use the code provided in the Supplementary Appendix to implement robust and linearized variance estimators in their own studies.

*Conclusion*

We alert researchers to the impact of the choice of the variance estimator when using the IPW method to handle attrition in cohort studies. We encourage the use of a robust or linearized variance estimator, and discourage the use of a naïve estimator of the variance.

**Conflict of interest**

None declared.

**References**

1. Little R, Rubin D. Statistical Analysis with Missing Data. New York: Wiley; 2002.

2. Hernán MA, Hernández-Díaz S, Robins JM. A Structural Approach to Selection Bias: Epidemiology. 2004 Sep;15(5):615–25.

3. Seaman SR, White IR. Review of inverse probability weighting for dealing with missing data. Stat Methods Med Res. 2013 Jun;22(3):278–95.

4. Weuve J, Tchetgen Tchetgen EJ, Glymour MM, Beck TL, Aggarwal NT, Wilson RS, et al. Accounting for Bias Due to Selective Attrition: The Example of Smoking and Cognitive Decline. Epidemiology. 2012 Jan;23(1):119–28.

5. Rabideau DJ, Nierenberg AA, Sylvia LG, Friedman ES, Bowden CL, Thase ME, et al. A novel application of the *Intent to Attend* assessment to reduce bias due to missing data in a randomized controlled clinical trial. Clin Trials J Soc Clin Trials. 2014 Aug;11(4):494–502.

6. Biele G, Gustavson K, Czajkowski NO, Nilsen RM, Reichborn-Kjennerud T, Magnus PM, et al. Bias from self selection and loss to follow-up in prospective cohort studies. Eur J Epidemiol. 2019 Oct;34(10):927–38.

7. Carpenter JR, Kenward MG, Vansteelandt S. A Comparison of Multiple Imputation and Doubly Robust Estimation for Analyses with Missing Data. J R Stat Soc Ser A Stat Soc. 2006;169(3):571–84.

8. Seaman SR, White IR, Copas AJ, Li L. Combining Multiple Imputation and Inverse-Probability Weighting. Biometrics. 2012 Mar;68(1):129–37.

9. Mansournia MA, Altman DG. Inverse probability weighting. BMJ. 2016 Jan 15;352:i189.

10. Robins JM, Rotnitzky A, Zhao LP. Estimation of Regression Coefficients When Some Regressors are not Always Observed. J Am Stat Assoc. 1994 Sep;89(427):846–66.

11. Beard JD, Hoppin JA, Richards M, Alavanja MCR, Blair A, Sandler DP, et al. Pesticide exposure and self-reported incident depression among wives in the Agricultural Health Study. Environ Res. 2013 Oct;126:31–42.

12. Rinsky JL, Richardson DB, Wing S, Beard JD, Alavanja M, Beane Freeman LE, et al. Assessing the Potential for Bias From Nonresponse to a Study Follow-up Interview: An Example From the Agricultural Health Study. Am J Epidemiol. 2017 Aug 15;186(4):395–404.

13. Deville J-C. Variance estimation for complex statistics and estimators: Linearization and residual techniques. Survey methodology. 1999;25(2):193–204.

14. Lunceford JK, Davidian M. Stratification and weighting via the propensity score in estimation of causal treatment effects: a comparative study. Stat Med. 2004 Oct 15;23(19):2937–60.

15. Lunceford JK. Stratification and weighting via the propensity score in estimation of causal treatment effects: a comparative study: Estimation of Causal Treatment Effects. Stat Med. 2017;36(14):2320.

16. Williamson EJ, Forbes A, White IR. Variance reduction in randomised trials by inverse probability weighting using the propensity score: Variance reduction in randomised trials by inverse probability weighting using the propensity score. Stat Med. 2014 Feb 28;33(5):721–37.

17. Hajage D, Chauvet G, Belin L, Lafourcade A, Tubach F, De Rycke Y. Closed-form variance estimator for weighted propensity score estimators with survival outcome. Biom J. 2018 Nov;60(6):1151–63.

18. Enders D, Engel S, Linder R, Pigeot I. Robust versus consistent variance estimators in marginal structural Cox models: Variance estimation in Cox MSMs. Stat Med. 2018 Oct 30;37(24):3455–70.

**Table 1. Response mechanism scenarios**

| Scenario | $\gamma_x$ | $\gamma_y$ | $\gamma_1, \gamma_2, \gamma_3, \gamma_4$ | Description |
|---|---|---|---|---|
| **MAR 1** | 0.0 | 0.0 | 0.1 | Response depending only on covariates |
| **MAR 2** | 0.2 | 0.0 | 0.1 | Response depending on covariates and exposure |
| **MAR 3** | 0.5 | 0.0 | 0.1 | Response depending on covariates and exposure |
| **MNAR 1** | 0.0 | 0.2 | 0.1 | Response depending on outcome and covariates |
| **MNAR 2** | 0.2 | 0.2 | 0.1 | Response depending on outcome, exposure, and covariates |
| **MNAR 3** | 0.5 | 0.2 | 0.1 | Response depending on outcome, exposure, and covariates |
| **MNAR 4** | 0.0 | 0.5 | 0.1 | Response depending on outcome and covariates |
| **MNAR 5** | 0.2 | 0.5 | 0.1 | Response depending on outcome, exposure, and covariates |
| **MNAR 6** | 0.5 | 0.5 | 0.1 | Response depending on outcome, exposure, and covariates |

$\gamma_k$: regression coefficients of the generated response models (logit$(p_i) = \gamma_0 + \gamma_y\, y_i + \gamma_x\, x_i + \gamma_1\, z_{1i} + \gamma_2\, z_{2i} + \gamma_3\, z_{3i} + \gamma_4\, z_{4i}$)

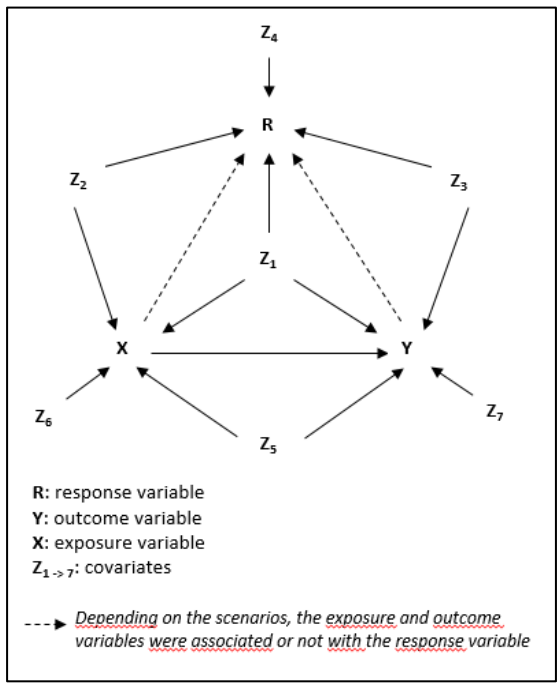

**Figure 1. Scheme of the data-generation model for the Monte-Carlo simulation**

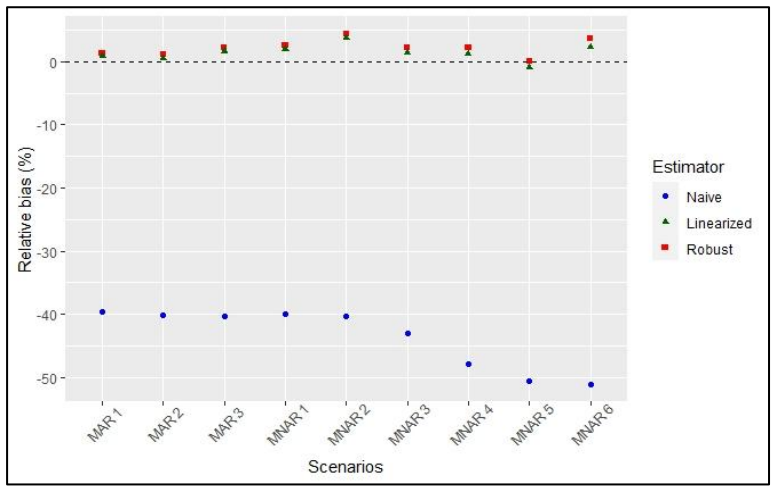

**Figure 2. Relative bias of the three variance estimators (naïve, robust, linearized) in the nine response mechanism scenarios**